\documentclass{elsart}
\usepackage{epsfig}

\def\astrobj#1#2{#2} 
\def\url#1{#1} 
\def\sfrac#1#2{#1/#2}

\begin{document}

\begin{frontmatter}

\title{The UV spectrum of nebulae}

\author{Fr\'ed\'eric \snm Zagury \thanksref{jsps}}

\address{Department of Astrophysics, Nagoya University,
\cty  Nagoya, 464-01 \cny Japan
\thanksref{email} }

\thanks[jsps]{Supported by the Japanese Society
for the Promotion of
Science,  grant ID No P97234.}
   \thanks[email]{E-mail: zagury@a.phys.nagoya-u.ac.jp}

\received{24 February 2000}
\accepted{14 April 2000}

\begin{abstract}
This paper presents an analysis of the UV spectrum of some nebulae with clearly
identified illuminating stars, all observed by the IUE satellite.

The data show remarkable properties of the UV spectrum of the nebulae. Each
spectrum is the product of the star spectrum and a linear function of
$1/\lambda$. There is no peculiar behaviour in the spectrums at 2200~\AA{}: 
no bump created in the spectrum of a nebula and no excess of scattering.
When moving away from the star, the surface brightness of a nebula decreases as
the inverse of the square of the angular distance to the star.

These results can logically be interpreted in terms of scattering of starlight.
They imply constant properties of the interstellar grains in the UV and in the
directions of space sampled by the nebulae, and probably a strong forward
scattering phase function. There is no evidence for any particular type of
grain which would specifically extinguish starlight at 2200~\AA{}.

Concerning the UV spectrum of a star, this may imply a revisal of the
traditional interpretation of the 2200~\AA{} bump.

\end{abstract}



\end{frontmatter}

\section{\label{Sec1}Introduction}
In the UV, as in the optical, the light which we receive from a nebula should
in great part be the light of the illuminating star scattered by dust grains in
the nebula. If no other process than scattering of starlight by the average 
population of interstellar grains is involved, the ratio 
of the nebula to the star spectrum depends on the optical depth of the nebula, 
on the angle $\varphi$ of
scattering, and on the albedo and phase function of interstellar grains.

The original idea of this work was to study the behavior of the spectrum of
nebulae in the 2200~\AA{} region. In this spectral region, the spectrum of
the stars is known to have a characteristic depression (the 2200~\AA{}
bump) when there is interstellar matter between the star and the observer. The
possible causes of the bump are reviewed by Bless~\& Savage \cite{bless72}.
Admittedly, the bump is an extinction feature (eventually a purely absorbing
one, Witt et al. \cite{witt92}) which arises due to a particular type of
interstellar grain.

Since a nebula placed in front of a star produces a bump in the spectrum of the
star, what will the bump spectral region of a nebula illuminated by a star at
close angular distance be? If the particles which extinguish starlight at
2200~\AA{} are present in the nebula, its spectrum should have an evident
feature at 2200~\AA{}. This feature will be an absorption one if the bump
carriers are purely absorbing particles. It will be an excess of emission if
the carriers scatter starlight.

Section \ref{data} presents the IUE observations used in this work. It consists
of the spectrums of a few well known nebulae, all illuminated by close bright
stars.

Section~\ref{analyse} to Section~\ref{analyse2}, is an analysis of the data. It
is a remarkable fact, and the most significant result of this analysis, that
all the nebulae have a spectrum which is exactly the product of the
illuminating star spectrum and a linear function of $1/\lambda$
(Section~\ref{analyse}). Some nebulae do not have a bump, and when they do have
one it is always proportional to the bump of the illuminating star. No bump is
created in the nebulae.

In Section~\ref{int}, the relation between the spectrum of a nebula and the
spectrum of the illuminating star is interpreted as scattering of starlight by
a medium with an optical depth which varies as $1/\lambda$.

A second result, the power law dependence of the surface brightness on the
angular distance to the star, is arrived at and discussed in
Sections~\ref{analyse2} and \ref{rdis}.
The large variations of the surface brightness level of a nebula with the
distance to the star may be an important argument in favor of a strong forward
scattering phase function in the UV (Section~\ref{rdis}). It implies that a considerable
amount of starlight is scattered in directions close to the star.

Different authors (e.g. A.N. Witt and collaborators) have found a strong
forward scattering phase function of the grains in the UV. These papers are
associated with models which also predict changes of the albedo and/or the
phase function with the wavelength. From analysis of the present data, I cannot
agree with these conclusions. Contrary to these studies, I find constant albedo
and phase function of the grains in all the UV, and perhaps up to the near
infrared, in all directions of space. In Section~\ref{grains}, a few of these
papers are briefly discussed along with the consequences the present paper have
on the properties of interstellar grains. It is suggested that there is no
excess of extinction in the spectrum of the stars at 2200~\AA{}.

An Appendix will review elementary properties of scattering and the notations
used in the paper.

\section{\label{Sec2}Data}
\label{data}
\subsection{\label{Subsec2.1}IUE data}
\label{iuedata}
All the observations used in this paper were made by the International
Ultraviolet Explorer. The IUE satellite and the detector system are described
by Boggess et al. \cite{boggess78a,boggess78b}. Recent and complete information
on the project and data reduction are available at the IUE website:
\url{http://iuearc.vilspa.esa.es}. The camera has two entrance apertures and 
two
dispersion modes. Observations can be made
through a large aperture ($L$), $10''\times 20''$ wide, or through a
small one, ($S$), $3''$ large in diameter. $S$-aperture observations of a star
generally need to be multiplied by a factor of $1.2$ to $2$ to match
$L$-aperture observations of the same star. The difference is attributed to the
difficulty of holding the star within the beam when using the small aperture.
Observations can be made in a low resolution mode of $\sim$7~\AA{} or with
a higher spectral resolution of 0.1--0.2~\AA{}. Three cameras: the LWR,
the LWP, and the SWP, have been used during the 20 year-long existence of the
satellite. The SWP camera covers the wavelength range 
[1150~\AA, 1980~\AA], while the LWR and the LWP cameras cover the 
[1850~\AA, 3350~\AA] range. The LWR camera was the `standard' long
wavelength camera from the beginning of the mission up until October 1983. A
problem with the camera's electronic led to a switch to the LWP. The total
range in wavelengths is from 1150~\AA{} to 3350~\AA. The data at
both boundaries should be carefully considered.

An IUE spectrum gives the
power per unit surface and per wavelength
received by the satellite in the U.V. from the direction of a celestial object. 
A typical extracted spectrum has the
$y$-coordinate in absolute flux units (erg/cm$^2$/s/\AA) and the $x$-coordinate
in~\AA. 
The sensitivity of the observations is limited to a few $10^{-15}$ erg/s/cm$^2$/\AA. 

The IUE data cannot be used straight away. A transfer function, which is to be
applied to the raw data, and which considerably modifies the original data
files, is provided by the IUE data reduction team.

I have used the .MLXO and .MXHI files which were up until recently available at
the IUE website. Those files are the final products of the
processing from raw data to final spectrums. Following IUE
recommendations, the files were read by the IDL software readmx.pro procedure
provided in the IUE package NEWSIPS for IDL.  

\subsection{\label{Subsec2.2}Data used in this paper}
\label{datapap}
\begin{table*}
    \[
\begin{tabular}{|c|c|c|c|c|c|c|c|c|}
\hline 
Name & $\alpha_{1950}$ &$\delta_{1950}$& star &$E$&$d_{\theta}$&
$\left(\sfrac{F_N}{F_{\star}}\right)_{m}$& 
$R^1$& $R\,\theta^2$
\\
&h m s & $\,\,^{\circ}\,\,\,\,\,\,\prime\,\,\,\,\prime\prime$ &&& $\prime\prime$ & $\times 10^{-3}$&&$\times
10^{-3}$ \\
 \hline
\astrobj{17 tau}{17$\tau$} nebula               & $03\,41\,55$   & $+23\,57\,10$ & \astrobj{HD 23302}{HD 23302}    & 0.047 & 18  & 1.81           & 0.56  & 2.93  \\
\astrobj{ori nebula}{Orion nebula} 1                & $05\,38\,18$   & $-02\,01\,05$ & \astrobj{HD 37742}{HD 37742}    & 0.07  & 193 & 0.0154         & 0.57  & 2.86  \\
\astrobj{ori nebula}{Orion nebula} 2                & $05\,38\,22$   & $-02\,03\,05$ & \astrobj{HD 37742}{HD 37742}    & 0.07  & 326 & 0.0048         & 0.51  & 2.56  \\
\astrobj{ori nebula}{Orion nebula} 3                & $05\,38\,34$   & $-02\,09\,05$ & \astrobj{HD 37742}{HD 37742}    & 0.07  & 727 & $8.5 \times 10^{-4}$ & 0.45  & 2.25  \\
\astrobj{merope nebula}{Merope nebula} 1               & $03\,43\,21$   & $+23\,46\,39$ & \astrobj{HD 23480}{HD 23480}    & 0.089 & 60  & 0.37           & 1.32  & 6.64  \\
\astrobj{merope nebula}{Merope nebula} 2               & $03\,43\,21$   & $+23\,47\,19$ & \astrobj{HD 23480}{HD 23480}    & 0.089 & 20  & 3.82           & 1.53  & 7.64  \\
\astrobj{20 tau}{20$\tau$} nebula 1             & $03\,42\,50$   & $+24\,12\,26$ & \astrobj{HD 23408}{HD 23408}    & 0.041 & 22  & 1.98           & 0.96  & 4.79  \\
\astrobj{NGC 2023}{NGC 2023}  & $05\,39\,06$   & $-02\,17\,24$ & \astrobj{HD 37903}{HD 37903}    & 0.32  & 26  & 14.98          & 10.12 & 50.63 \\
\astrobj{IC 435}{IC 435}      & $05\,40\,30$   & $-02\,19\,48$ & \astrobj{HD 38087}{HD 38087}    & 0.25  & 17  & 7.98           & 2.31  & 11.53 \\
\astrobj{20 tau}{20$\tau$} nebula 2             & $03\,42\,51$   & $+24\,12\,32$ & \astrobj{HD 23408}{HD 23408}    & 0.041 & 15  & 3.99           & 0.90  & 4.49  \\
\astrobj{NGC 7023}{NGC 7023}  & $21\,00\,56$   & $+67\,57\,49$ & \astrobj{HD 200775}{HD 200775}  & 0.55  & 20  & 1.07           & 0.44  & 2.2   \\
\astrobj{CED 201}{CED 201}    & $22\,12\,17$   & $+70\,00\,25$ & \astrobj{BD +69 1231}{BD +69 1231}  & 0.16  & 15  & 10.98          & 4.39  & 12.35 \\
\hline
\end{tabular}
\]
\caption[]{Main parameters for the nebulae presented in the paper. Nebulae are listed
by increasing optical depth. The table gives the adopted name of the nebula;
coordinates of IUE nebula observations; its illuminating star;
$E=E(B-V)$ in the star direction;
angular distance to the star; 
maximum observed ratio of nebula to star fluxes,
$\left(\sfrac{F_N}{F_{\star}}\right)_{m}$; 
$\left(\sfrac{F_N}{F_{\star}}\right)_{m}$ rescaled to a distance of $1''$: 
$R^1=\left(\sfrac{F_N}{F_{\star}}\right)_{m}^{1''}$; and  $R=
 \left(\sfrac{F_N}{F_{\star}}\right)_{m}^{1''}/200''^2$.}
\label{tbl:neb}
\label{Table-1}
\end{table*}
I have used the spectrums of nebulae (IUE class~73) and of stars. For each
object, the spectrum presented in the paper is an average of the best
observations of the object. I sometimes used high dispersion spectrums and
decreased the resolution by using a median filter.

Observations of nebulae with clearly identified illuminating sources are not so
frequent.
Most were studied although they are not all presented here. The results of the
paper hold for the other nebulae as well.

The main properties of the nebulae are summarized in Table \ref{tbl:neb}. I
found it unnecessary to specify the IUE observations' name of the objects
listed, since they can easily be looked at or retrieved from the IUE website. A
reference list of the IUE images is available upon request.

All nebulae were observed at different distances to the star. Different
observations of a nebula with the same camera are often proportional when the
position of the observations are close. This proportion is justified by
the proximity of the observations and indicates equal optical depths and
similar mediums. They may differ when the positions are too far apart due to
modifications of the nature of the interstellar medium. When different
positions to the star were exactly proportional, I chose to present either the
best spectrum or the average one, so as to reduce the noise. Regions
illuminated by the same star, with spectrums which cannot be superimposed, were
considered as separate nebulae.

The spectrums of the same object taken with the SWP camera or with one of the
long wavelength range cameras usually join in the common spectral region
 ($\sim$1900--2000 \AA). 
 A slight gap between the long and the short
wavelength spectrums
sometimes happens and has been attributed either to a calibration
problem or to a problem of precision in the pointing. In this case a correcting
factor was applied to the long wavelengths' range spectrum to ensure continuity
from one range of wavelengths to another. This scaling has no effect on the
relations which will be obtained between the spectrums of a
nebula and that of its illuminating star since these relations apply
individually to each camera.

\section{\label{Sec3}Comparison of the spectrums of the nebulae with that of the stars}
\label{analyse}
\subsection{\label{Subsec3.1}Nebulae with no bump}
\label{neb0}

\begin{figure}
\resizebox{\columnwidth}{!}{\includegraphics{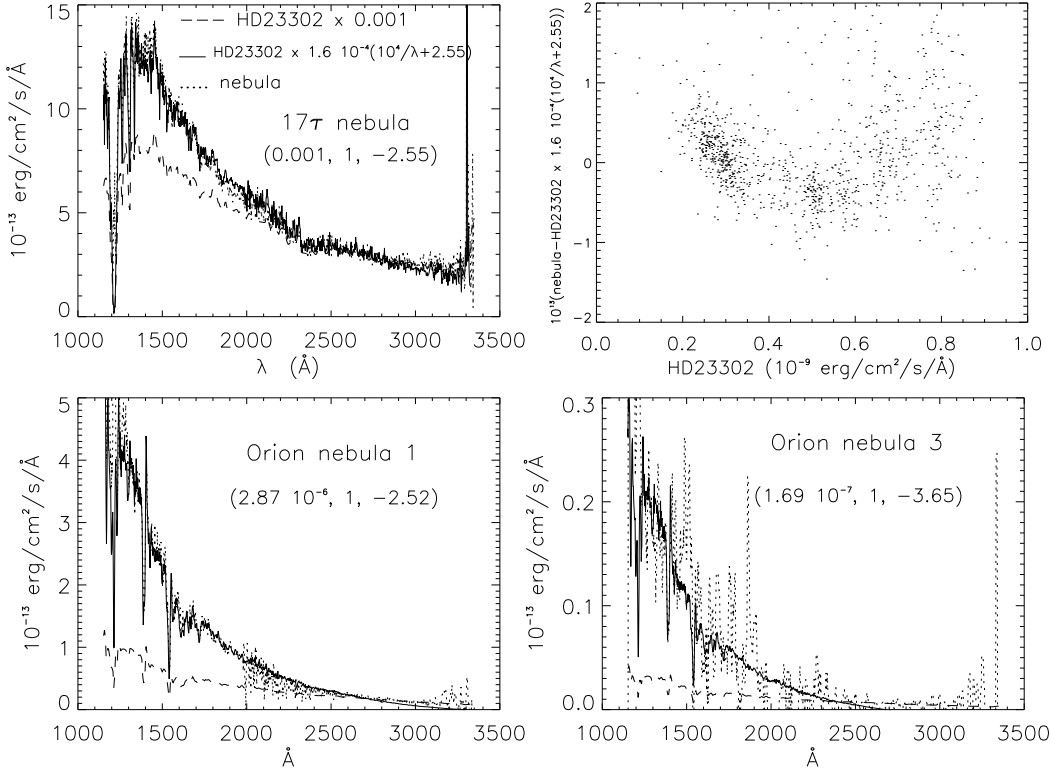}}
\caption[]{\emph{Top left}: The spectrum of the \astrobj{17 tau}{$17\tau$} nebula, 
in dots, is overlaid by the
product of the star spectrum and a linear function in $1/\lambda$. Underneath,
the star spectrum (in dashes) was scaled to the nebula's in the long wavelength
range. \emph{Top right}: Difference between the star-fit to the nebula 
and the nebula spectrums, plotted versus the star spectrum. There is no 
residual correlation. \emph{Down}: Spectrums of \astrobj{ori nebula}{Orion nebula}e~1 and 3 and of
the illuminating star \astrobj{HD 37742}{HD 37742}. 
Dash, dot and solid lines have the same meaning
as for the \astrobj{17 tau}{${17}\tau$} nebula. The signification of the numbers in parenthesis
is given in Section~\ref{rto}.}
\label{17tau}
\label{Figure-1}
\end{figure}

Fig.~\ref{17tau}, left, plots the spectrum of a nebula and its illuminating
star, \astrobj{HD 23302}{HD 23302} (\astrobj{17 tau}{$17\tau$}), 
as a function of $\lambda$. The two bottom plots are
two positions observed by the IUE satellite in the \astrobj{ori nebula}{Orion nebula}. In both cases,
the illumination is due to \astrobj{HD 37742}{HD 37742}, a few arcminutes apart. An intermediate
position, \astrobj{ori nebula}{Orion nebula} 2, not represented here, was also observed. The spectrum
of the \astrobj{ori nebula}{Orion nebula} 2 
matches the \astrobj{ori nebula}{Orion nebula}~1 spectrum when multiplied by 3.

Stars and nebulae spectrums are not proportional, but Fig.~\ref{17tau} shows
that the spectrum of a nebula is exactly overlaid by the spectrum of the
illuminating star, after multiplication by a linear function of $1/\lambda$.

It is worth noticing that no bump is created in the nebulae.

\subsection{\label{Subsec3.2}General relation between the UV spectrums of a nebula
and that of the illuminating star}
\label{rto}
\begin{figure*}[p]
\resizebox{\columnwidth}{!}{\includegraphics{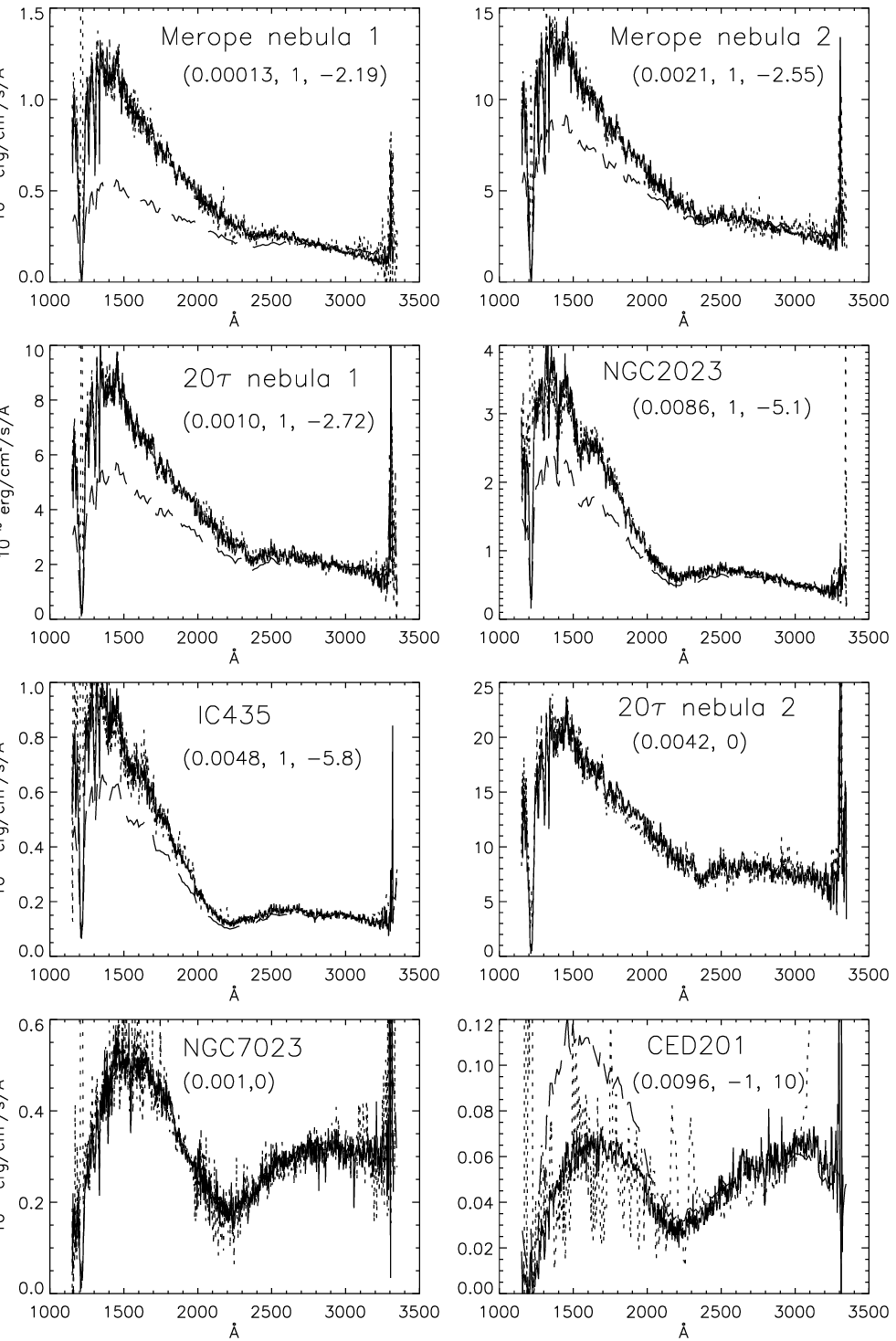}}
\caption[]{Spectrums of 8
nebulae and of their illuminating star. For each plot the nebula spectrum is in
dots, the star spectrum (smoothed and scaled by $k$ to fit the nebula spectrum
at long wavelengths) in dashes, and the star spectrum fitted to the nebula's in
solid line. The coefficients are $(k, a, b/a)$, $k$ defined above and $(a, b)$
from Eq.~(\ref{transfert}). When $a=0$, $b=\infty$ is omitted. The nebulae
are presented by order of increasing average optical depth from left to right
and top to bottom.}.
\label{nebfig}
\label{Figure-2}
\end{figure*}

The preceding fit of the emission of a nebula by the product of the
illuminating star flux and a linear function of $1/\lambda$ can successfully be
repeated for all the nebulae observed by the IUE satellite, whether they have a
bump or not. Fig.~\ref{nebfig} shows the results for 8 nebulae and gives 3
coefficients, $(k,a, b/a)$.  $k$ is
the coefficient of the zero order fit, restricted to the long wavelength's
range, of the spectrum of the star to the nebula spectrum. The star spectrum,
after smoothing by a median filter for sake of clarity, is multiplied by $k$
and represented by the curve in dashes. The nebula spectrum is in dots. It is
fitted by the star spectrum multiplied by a linear function of $1/\lambda$, the
overlaid solid line curve. The relation between the spectrums can 
always be written as:
\begin{equation}
F_{{N}}\sim \alpha F_{\star} 
\left( a \, \frac{1 \,{\rm \mu m}} {\lambda} -b\right),
\label{transfert}
\end{equation}
where $F_{{N}}$ and $F_{\star}$ are the nebula and the star fluxes measured by
the IUE satellite. $\alpha$ is a coefficient which must depend on the albedo
and the phase function of interstellar grains and on the relative
position of the star, the nebula and the observer. $a$ can be 0, 1 or $-$1.

The relation which is found between the UV emission of a nebula and its
illuminating star should still apply to their UV fluxes corrected for
foreground extinction: the star and the nebula are close enough to be 
equally (in proportion of their flux)
affected by interstellar material situated between the star and the nebula on
the one hand and the observer on the other hand. Since they are proportional,
the bumps of a nebula and its illuminating star should be attributed to dust in
front of both the nebula and the star.
\section{\label{Sec4}Strength and variation of UV surface brightness of the nebulae
with distance to the star}
\label{analyse2}
The proportion of starlight which is scattered by a nebula is evaluated by the
$R=S/F_{\star}$ parameter. $R$ can theoretically be defined as the ratio of the
nebula surface brightness to the star flux, both measured at the same position
and corrected for foreground extinction. Slight or no reddening of the star's
light between the star and the scattering volume should also be assumed.

The maximum proportion of starlight which can be scattered in the
direction $\varphi$ is given by 
$R_{{max}}=S_{{max}}/F_{\star}$, where $S_{{max}}$ is the
maximum surface brightness the nebula can reach (corresponding to an optical
depth $\tau_{max}\sim 1-2$) at $\varphi$. $R_{{max}}$ depends solely on the angle of
scattering $\varphi$.

Column~5 of Table~\ref{tbl:neb} gives the non-dimensional quantity $R\theta^2$
for each of the observations. $R\theta^2$ was estimated by taking the 
maximum value (over the $[1000\rm\AA,3000\AA]$ IUE wavelength range) of $(F_{N}/F_{\star})(\theta^2 /A_{IUE})$, 
the ratio of the maximum surface brightness of the
nebula to the star flux, normalized to a distance of $\theta=1''$ to the star.
$A_{{IUE}}=200$ arcsec$^2$ is the surface of the IUE camera large aperture.
If, as will be suggested in Sect.~\ref{intopt}, the mean optical depth of 
the nebulae is close to $\tau_{max}$, $R\theta^2$ should be close to 
$R_{max}\theta^2$.

The range of values the observed maximum of $R\theta^2$ takes (last column of
Table~\ref{tbl:neb}) is quite narrow. Within the restrictions due to the margin
of error discussed in the following section, $R_{{max}}\theta^2$ can be taken to be
$\sim 3 \times 10^{-3}$, for $\theta$ between $\sim 10''$ and a few arcminutes. From
the present data, it is not possible to discern whether the observed 
maximum of $R\theta^2$ in Table~\ref{tbl:neb} is a
constant or varies with the nebula.

Table~\ref{tbl:neb} suggests a power law dependence of the nebulae surface
brightness with angular distance to the illuminating star. For the observations
of nebulae in the same region (e.g. 
\astrobj{ori nebula}{Orion}, 
\astrobj{merope nebula}{Merope} and 
\astrobj{20 tau}{$20\tau$} nebulae), 
illuminated by the same star, $R\theta^2$ is remarkably constant.

When different positions of the same nebula are available and
proportional, the data show that there
are strong variations of the brightness with angular distance to the star but
$R\theta^2$ stays relatively constant.

\subsection{\label{Subsec4.1}Sources of error on the evaluation of the $R$ coefficient}
\label{rerr}
Differences between the observed values of $R$, reported in the
Table~\ref{tbl:neb}, and its theoretical definition can arise from a mistaken
evaluation of either the nebula surface brightness or the unreddened flux of
the star.

A major difficulty when estimating the exact surface brightness is the width of
the beam of the observations, which in many cases has dimensions comparable to
the distance to the star. It should be reliable when the dimensions of the beam
are small compared to the distance to the star but errors may arise when the
observation is made at close distance to the star (a few $10''$). In the latter
case, the variation of the nebula surface brightness between the two ends
(facing and opposite the star) of the $L$-aperture, and its orientation, have
to be taken into account. The margin of error can be estimated by comparing the
brightnesses obtained from observations at the same position with the
$L$-aperture on the one hand, and from the $S$-aperture on the other hand.

The expected ratio is 7. Few nebulae were observed with the $L$ and the $S$
aperture. The ratio of the brightnesses in these observations ranges from 7, as expected, to 100.
Although the pointing may explain part of this large factor, the exact surface brightness of a nebula, estimated by 
dividing the brightness of the nebula ($F_N$ in Eq.~(\ref{transfert})) 
by the solid angle intercepted by the $L$--aperture, may be overestimated by a factor of 10, 
for positions at less than a few
$10''$ to the star.

The second uncertainty in the evaluation of $R$ comes from the radiation field.

If the starlight is extinguished between the star and the scattering volume and if
material in front of the star does not compensate for this extinction, the
observed value of $R$ is superior to the exact one.
Conversely, if there is more extinction between the star and the observer than
between the nebula and the observer, the observed value of $R$ will be an
inflated estimate of the exact one.
Table~\ref{tbl:neb} shows a tendency for the observed maximums of $R\theta^2$ to increase with the
reddening of the star. This increase may in part be explained by extinction of
starlight in the vicinity of the star.

\section{\label{Sec5}Interpretation of Equation~(\ref{transfert})}
\label{int}

\subsection{\label{Subsec5.1}The variations of the brightness of a nebula with optical depth}
\label{intopt}
The stars \astrobj{17 tau}{$17\tau$} 
and \astrobj{HD 37742}{HD 37742} have no bump and are slightly reddened
($E(B-V)=0.07$, Table~\ref{tbl:neb}). The nebulae emissions, which have no
peculiarities at 2200~\AA, can safely be attributed to scattering.
The linear transformation which permits passage from the star to the nebula
spectrum is to be interpreted as the slope of the surface brightness versus the
optical depth curve (Fig.~\ref{bumpfig}) in the optical depth range of the
spectrum.

The same should hold for nebulae with bump. Eq. (\ref{transfert}) expresses
that around the mean optical depth $\tau_{{mean}}$ of the nebula and for the
wavelength range under consideration, $F_N$ variations with
$\tau_{\lambda}$ can be approximated using the slope of $S(\tau )$
(Fig.~\ref{bumpfig}) at $\tau_{{mean}}$. If $a=1$, $F_{{N}}$ grows with
$\tau_{\lambda}$ and $\tau_{{mean}}$ 
must be less than $\tau_{{max}}$. If $a=0$,
$\tau_{\lambda}\sim\tau_{{max}}$, the spectrums are close to being proportionate.
Last, if $a=-1$, $\tau_{{mean}}>\tau_{{max}}$.

The $b$ coefficient has a remarkable property. $ a(1{\rm \mu m}/\lambda) -b$ is
the only source of $F_N$ variations with $\lambda$ and must be proportional to
the proportion of scattered starlight (restricted to $\tau_{\lambda}$-values
close to $\tau_{{mean}}$), the $y$-axis of Fig.~\ref{bumpfig}. Hence, $b/a$ will
always correspond to the point where the slope crosses the abscissa, with $'x'$
scaled to be $1/\lambda$. If $a=1$; the slope is positive,
$\tau_{{mean}}<\tau_{{max}}$, 
and $\tau_{{mean}}$ increases when $b\,(<0)$ decreases
(Fig.~\ref{bumpfig}). If $a=-1$; the slope is negative,
$\tau_{{mean}}>\tau_{{max}}$, 
and $\tau_{{mean}}$ increases with $b/a\,(>0)$, if $\tau_{mean}$ is not 
too large.
The $b/a$ coefficient provides a way
to classify nebulae according to their optical depths.

The spectrums of Fig.~\ref{nebfig} are plotted by increasing optical depths,
according to the considerations of the previous paragraph. Two nebulae only,
\astrobj{20 tau}{20$\tau$} nebula~2 and \astrobj{NGC 7023}{NGC 7023} 
have their spectrums in exact proportion to the
spectrum of the star. These nebulae have an optical depth $\tau_{{mean}}$ equal
to $\tau_{{max}}$. Their surface brightness, as a function of the optical depth,
is at its maximum.

There are striking relations between the optical depth and the changes in the
shape of the UV spectrum of Fig.~\ref{nebfig}. Increasing the optical depth
of a nebula will soften the slope of the rise between 2200  \AA{} and 
1300  \AA. The effect can be observed by comparing the spectrums of the stars
and the nebulae. Increasing the optical depth also attenuates the FUV emission.
Both phenomenae are the consequences of the increase of absorption toward
shorter wavelengths, and contribute to the emphasis of the bump's visual
impression. It is particularly evident for \astrobj{CED 201}{CED 201}.

Most of the nebulae spectrums are close to being in proportion with the
illuminating star's spectrum: for each nebula the zero order fit to the
spectrum of the star is in reasonable agreement with the spectrum of the star.
It can be interpreted as 
$\tau_{{mean}}$ values which are close to $\tau_{{max}}$.
$R$, calculated as indicated in Section~\ref{analyse2}, must be close to
$R_{{max}}$, defined in the same section.

\subsection{\label{Subsec5.2}The albedo and the phase function}
\label{intalb}

The interpretation which is given of the Eq. (\ref{transfert}) involves only
the change in optical depth with wavelength. There is no evidence or need for a
variation of the albedo or of the phase function of interstellar grains in the
wavelength range [1000, 3000] \AA{} of the IUE observations.

There is also no evidence for variations of the properties of the grains with
the direction of space.

\section{\label{Sec6}The phase function of interstellar grains in the UV}
\label{rdis}
The analysis of Eq. (\ref{transfert}), has focused on the variation of the
brightness of a nebula with wavelength, which depends on variations of the
optical depth only.

The effect of the phase function of interstellar grains on the brightness of a
nebula will be more evident through the comparison of the brightnesses of
different observations because each of these will correspond to different
angles of scattering, $\varphi$ in Fig.~\ref{schema}. IUE observations are
particularly useful in this respect since they provide observations of the same
nebula at different angular distances to the star and observations of nebulae
illuminated by different stars.

Assume an isotropic phase function. $F_N$ will depend on the radiation
field at the position of the scattering volume and not on $\varphi$. The large
variations of the brightness with the distance to the star ($R\propto
\theta^{-2}$, Section~\ref{analyse2}) will be justified only if $\theta$ is
proportional to the distance $d$ of the scattering volume to the star. It
implies that the nebulae are all approximately at the same distance as the star
from the observer: $d_0\sim 0$ and $d=D\theta$. The surface brightness of the
nebula is:
\begin{equation}
S=\frac{L_{\star}}{4\pi d^2} \, {Sca},
\label{eq:sbiso}
\end{equation}
with $ L_{\star}$ the luminosity of the star and $Sca$ the proportion of
scattered starlight by a medium of optical depth $\tau$ in
direction $\varphi$. For $\tau \sim \tau_{{max}}$, the simple model developed in
Zagury, Boulanger and Banchet \cite{zagury99}, shows that $Sca_{{max}}$ should be
at least $\omega e^{-1}/(4\pi)$. $R_{{max}}\theta^2$ will be found to be:
\[
R_{{max}}\theta^2  =  
\frac{S_{{max}}}{\sfrac{L_{\star}}{4\pi D^2}} \, \theta^2
>  \frac{\omega e^{-1}}{4\pi} ,
\]
\begin{equation}
R_{{max}}\theta^2 > 0.03\omega.
\label{eq:riso}
\end{equation}
Even though there is some uncertainty in the observed values of
$R_{{max}}\theta^2$ ($\sim$0.003, Section~\ref{analyse2}), isotropic scattering
probably implies an albedo of at most $0.1$.

This very low albedo, and the systematic configuration in which all the nebulae
need to be to justify isotropic scattering, are improbable. A more likely
explanation for the variation of the surface brightness with the angular
distance to the star can be attributed to variations of $\varphi$ rather than
variations of $d$. The quick decrease of the surface brightness with $\theta$,
i.e. with $\varphi$, indicates a strong forward scattering phase function.
Assuming a strong forward scattering phase function, the scattering volumes are
likely to be in front of the stars, in which case we have $d\sim d_0$
(Fig.~\ref{schema}).

Three independent factors will modify $S_{{max}}$: the star luminosity,
$L_{\star}$, the distance to the star, $d$, and the angle of scattering,
$\varphi$. The surface brightness of a nebula is proportionate to $L_{\star}$
and to $d^{-2}$. The remaining dependence on $\varphi$ will be defined by the
function $h(\varphi)=4\pi d^2/ L_{\star}\,S_{{max}}$. Then:
\begin{equation}
R_{{max}} =   \frac{\sfrac{h(\varphi)L_{\star}}{d^2}} {\sfrac{L_{\star}}{D^2}}
 =  
h(\varphi) \, \frac{\varphi^2}{\theta^2}\, .
\label{eq:hphi}
\end{equation}
The dependence of $R_{max}$ on $\theta^2$ in 
equation~\ref{eq:hphi} was e.g. verified by the observations 
(section~\ref{analyse2}). 
Concerning the observations for which $R_{max}\theta^2$ 
is nearly constant ($R_{max}\theta^2=c\sim 3\,10^{-3}$),
we have: $h(\varphi)=c/\varphi^2$.
Since $h$ does not depend on $\theta$ this relation must be true for the range 
of $\varphi$--values considered by the observations.
It suggests that the
differences between the values of $c=R\theta^2$ in Table~\ref{tbl:neb} is due
to errors on the estimate of either $R$ (Section~\ref{Subsec4.1}) or $\theta$.

Hence, the maximum surface brightness a nebula can reach in the direction $\varphi$ from the
direction of the star varies as:
\begin{equation}
S_{{max}}=\frac{c}{\varphi^2}\frac{L_{\star}}{4\pi d^2} \, .
\label{eq:sb}
\end{equation}
Eq. (\ref{eq:sb}) applies to $\varphi=\theta D/d$ values between a few
$10''$ (or less) and a few $10'$.

\section{\label{Sec7}Further implications}\label{grains}

\subsection{\label{Subsec7.1}The UV spectrum of the stars}\label{bump}

An important consequence of the relation found between the spectrum of 
the nebulae and the spectrum of the illuminating stars concerns the 2200~\AA{} bump.

In none of the UV observations of the nebulae presented here is there evidence
for the presence of the 2200~\AA{} carriers. If starlight is not affected
by the presence of the bump carriers when the nebula is on the side of its
illuminating star, why should it be affected when the nebula is in front of the
star?

From Eq. (\ref{eq:sb}) it is clear  that considerable starlight
can be scattered into the beam of the telescope at close angular distance to a
star. If $\varphi_{{min}}$ is the smallest angle down to which
Eq. (\ref{eq:sb}) holds, and if we assume $R_{{max}}$ to be constant for
$\varphi<\varphi_{{min}}=\theta_{{min}}D/d$, 
the proportion of starlight scattered
into the beam of semi angle $\theta_0$ can reach:
\begin{eqnarray}
    \int_0^{\theta_0}2\pi R_{max}\theta{\rm d}\theta
    &=&\int_0^{\theta_{min}}\frac{2\pi c\theta}{\theta_{min}^2}{\rm 
    d}\theta + 
     \int_{\theta_{min}}^{\theta_0}\frac{2\pi c}{\theta}{\rm d}\theta
    \nonumber\\
    &=& \pi c
    (1+2\ln(\frac{\theta_0}{\varphi_{min}}\frac{D}{d}))
    \label{eq:bump0} \\
     &\sim& 10^{-2}
    (1+2\ln(\frac{\theta_0}{\varphi_{min}}\frac{D}{d}))
\label{eq:bump}
\end{eqnarray}
It can easily be a few percent of the direct starlight corrected for
extinction. The proportion of scattered starlight into the beam will be even
more important when compared to the direct starlight $F_{\star}$ received by the
telescope, since the latter is extinguished.

The implication may be that there are no bump carriers. It was noted by
Bless~\& Savage \cite{bless72} (see also Cardelli~\& Clayton 
\cite{cardelli88} pp.~519 and 520)
that the bump will also be explained if enough
starlight scattered in the beam of the observation was added to the direct
starlight. This possibility was ruled out (Bless~\& Savage \cite{bless72}) on the basis of
arguments which do not hold if grains strongly scatter light in the forward
direction.

\subsection{Interstellar grains properties}\label{grainsalb}

\subsubsection{ Albedo and phase function}\label{alb}

Contrary to Witt \cite{witt85}, who finds variations of grain properties in one
direction to another, I find homogenous properties of interstellar grains in
all directions and constant albedo and phase function in the UV.

Witt~\&~Bohlin \cite{witt86} claim the discovery of enhanced scattering
emission in reflection nebulae associated with the 2175 \AA{} extinction
band but no such irregularities can be observed in any of the nebulae presented
here. The spectrums of \astrobj{CED 201}{CED 201} and 
\astrobj{IC 435}{IC 435} presented in Witt~\&~Bohlin are close
to the raw data which may mean they have not been treated according to 
IUE most recent recommendations. 
Consequently, large differences exist between the nebulae
spectrum in Witt~\&~Bohlin and the ones presented at the IUE website and
reproduced here. The long wavelength range observation of 
\astrobj{CED 201}{CED 201} used by the
authors, LWR~17725, on which most of their discussion relies, is classified
`$a/n$' or `abnormal reading' and `non standard acquisition'
at INES Archive Data Mirror.

The spectrums of \astrobj{IC 435}{IC 435} and 
\astrobj{CED 201}{CED 201} presented in Fig.~\ref{nebfig} are an
average of the best observations of the nebulae, most of which were not
available at the time Witt~\&~Bohlin did their work. The spectrums have no
unique features, except for the large optical depth met in \astrobj{CED 201}{CED 201}.
 It is most
probable that Witt~\&~Bohlin's analysis as well as their conclusions are
questionable.

Another study, the Calzetti et al. \cite{calzetti95} paper on 
\astrobj{IC 435}{IC 435}, uses the
same data as that found in the present paper.
The authors have followed the Witt~\& Bohlin analysis and attribute the
variation of $\log (S_{{N}}/F_{\star})$ with wavelengths to changes in the albedo
and/or phase function of the grains. They didn't search for a relation between
the star flux, the nebula surface brightness and $1/\lambda$.

\subsubsection{Power law dependance of the surface brightness on $\theta$}\label{ph}
\begin{figure}
\resizebox{\columnwidth}{!}{\includegraphics{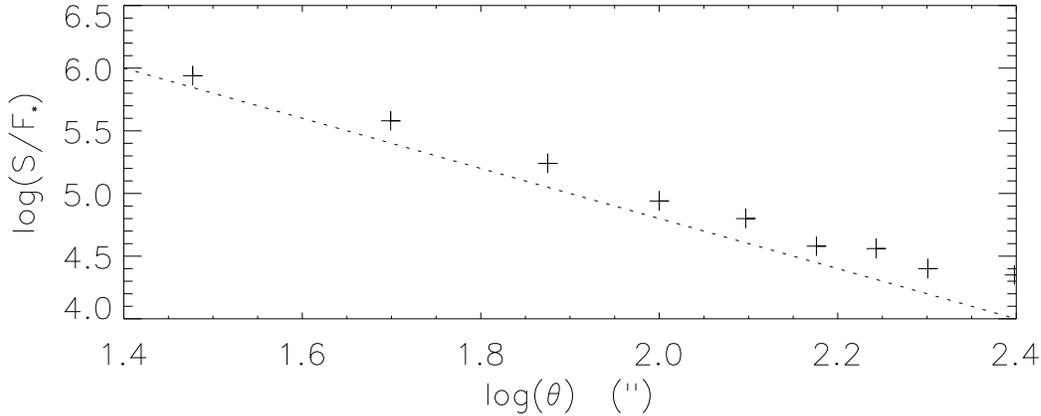}} 
\caption[]{$\log
(S/F_{\star})$ versus $\log\theta$ from the observations of Witt~\& Cottrell
\cite{witt80}, table~II, field North. The dotted line has a slope of $-2$. The
same dependence of $S/F_{\star}$ on $\theta^{-2}$ is found for the other fields
(W, S, E).}
\label{witt80fig}
\label{Figure-3}
\end{figure}
The power law dependence of the surface brightness of 
\astrobj{NGC 7023}{NGC 7023} with distance to
the star was investigated from the optical to the UV by Witt~\& 
Cottrell~\cite{witt80} and Witt et al.~\cite{witt92}.

The power law which is found in these two papers has an exponent $\sim$1.5 in
absolute value, smaller than the one found in this paper. This exponent is not
the one which can be deduced from the raw data since the authors try to include
a correction for the width of the aperture, which introduces an error on the
surface brightness estimate at close angular distance to the star.

The raw data are not given in Witt et al.~\cite{witt92}. The observational
results given in Witt~\& Cottrell~\cite{witt80} for the optical wavelengths
give an exponent of $2$ (Fig.~\ref{witt80fig}). Other observational data, for
instance the Witt~\& Schild~\cite{witt86} study of fifteen reflection nebulae, give a similar
exponent (Zagury, submitted).

Precise comparison of the different papers cannot be broached here. An exponent
of 2 is logical within the present analysis. The exponent of 1.5 found in
Witt et al.~\cite{witt92} is also justifed by the model developed in their
paper.

Another idea which emerges from the comparison of the data in the different
papers mentioned in this section, is that the UV and optical wavelengths for
which $R\theta^2$ can be calculated give similar values. If true, it may imply
similar properties (albedo and phase function) of interstellar grains from the near infrared to the far UV.

\section{\label{Sec8}Conclusion}
Comparison of the UV spectrum of several nebulae observed by the IUE satellite
to the spectrum of the illuminating stars shows that the spectrum of a nebula is always the
product of two terms: the flux of the star $F_{\star}$ and a linear function of
$1/\lambda$. This function was interpreted as the slope of the surface
brightness $S$ of the nebula versus optical depth at the optical depth $\tau_{{mean}}$ of
the nebula.
The variation of $R=S/F_{\star}$ with wavelength, and of its maximum with the
angular distance $\theta$ to the star, provides additional informations on the
properties of interstellar dust and on the nebulae.

The linear $1/\lambda$ function gives an estimate of $\tau_{{mean}}$, since the
slope of $S(\tau)$ decreases with $\tau$. It allows a classification of the
nebulae according to their mean optical depth. For all the nebulae of the
sample, $\tau_{{mean}}$ should be close to $\tau_{{max}}$, the optical depth at
which maximum surface brightness is reached. This is in conformity with the
high brightnesses of the nebulae.

The important variations of the ratio $R=S/F_{\star}$ with the angular distance $\theta$ to
the star is explained if the interstellar grains have a strong forward
scattering phase function in the UV.

In some observations the maximum value of $R$, $R_{{max}}$
varies as $1/\theta^2$. 
This result is justified if the phase function is so that $R_{{max}}$
 is proportional to $1/\varphi^2$, where $\varphi$ is the angle of scattering.

No additional absorption or other particular behaviour at 2200~\AA{} was
noticed in the spectrum of the nebulae. The absence of bump carriers in the
nebulae calls to question the interpretation of the 2200~\AA{} bump in the
spectrum of the stars. 
When a star is observed behind a nebula
a considerable amount of starlight scattered into the beam can be 
added to the direct starlight and may explain the bump.

The understanding and interpretation we have reached of the UV spectrum 
of nebulae applies similarly to different regions of the interstellar 
medium.
It indicates that the UV properties (albedo and phase function) of interstellar 
grains are identical in all directions of space. 
These properties may be the same from the near infrared to the far UV.

\appendix
\section{\label{Sec9}Appendix: Variations of a nebula surface brightness with optical depth}
\begin{figure}
\resizebox{0.5\columnwidth}{!}{\includegraphics{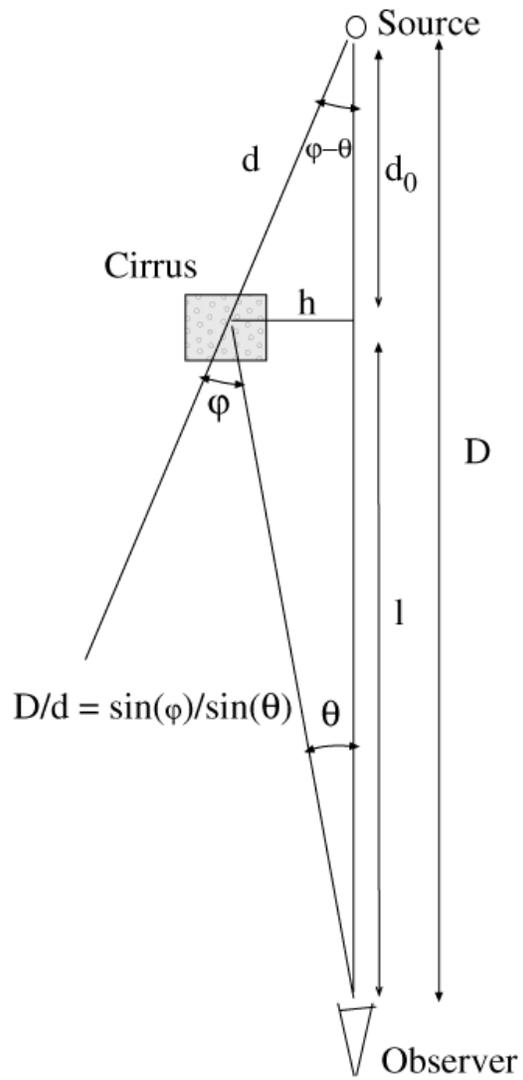}}
\caption[]{Representation of the angles employed in the text.}
\label{schema}
\label{Figure-4}
\end{figure}
\begin{figure*}
\resizebox{\columnwidth}{!}{\includegraphics{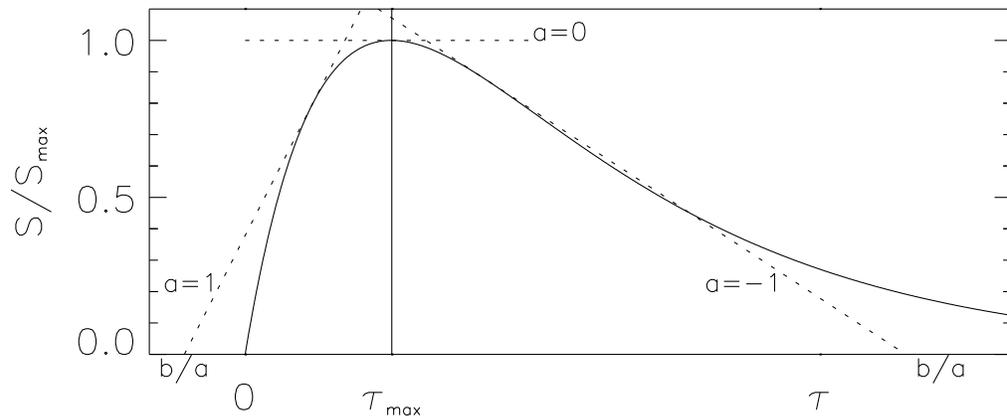}}
\caption[]{Surface brightness (scaled to $1$ at maximum) versus optical depth.}
\label{bumpfig}
\label{Figure-5}
\end{figure*}
Elementary properties of scattering used in this paper are summarized by the
plot of Fig.~\ref{bumpfig}. The plot represents surface brightness
variations with optical depth for a cloud illuminated by a far away source (the
source of light comes from one direction).

The $x$-axis of the plot is the optical depth, related to the quantity of
interstellar grains photons have crossed on their way to the telescope. The
optical depth depends on wavelength and on the column density along the path
the photons have followed. Optical depth and column density are proportional.
Optical depth is a linear function of $1/\lambda$. Thus, depending on the kind
of observation which is involved, spectroscopy or photometry, $\tau$ will vary
from a point to another according to the wavelength or to the quantity of
matter on the photons' way. The quantity of matter on the photons' way depends
on the angle made by the observer, the source radiation field and the nebula.
Only for directions close to the illuminating source can $\tau$ be assimilated
to the cloud column density.

The $y$-axis is the nebula surface brightness scaled to 1 at maximum surface
brightness. At very small $\tau$, variations of $S$ and $\tau$ are
proportional. Absorption will quickly diminish the slope d$S$/d$\tau$. $S$
will have a maximum at a value $\tau=\tau_{{max}}$ and decreases for
$\tau>\tau_{{max}}$. For any given column density, there is a wavelength at which
$\tau=\tau_{{max}}$ is reached and reciprocally. Hence there is a correspondence
between wavelengths and column densities which allow to predict at which
wavelength a cloud will preferably be observed and the possible range of column
densities of a cloud bright in a certain wavelengths range.

For a constant albedo, $S_{{max}}=S(\tau_{{max}})$ should depend solely on the
phase function, i.e. only on the angle $\varphi$ of the source radiation field
direction with the direction of the scattering volume and the observer.

\ack
I thank Professor Y.~Fukui for the support and freedom I took benefit from
during two years in Japan and without which this work would not have been
possible. 
I received constant help and a warm welcome from all the Radio Astronomy group of Nagoya
University. A special thank to Dr~Onishi and to Dr~Mizuno.

Dr E.~Soleno, from ESA, introduced me to the use of the IUE data and 
answered many questions. I am also indebted to
Stacey Young for having reviewed many times the english of the manuscript.


\begin{thebibliography}{}

   \bibitem{bless72} Bless R.C., Savage
B.D, 1972, ApJ, 171, 293

   \bibitem{boggess78a} Boggess A., et al., 1978,
Nature, 275, 372

   \bibitem{boggess78b} Boggess A., et al., 1978, Nature,
275, 377

   \bibitem{calzetti95} Calzetti D., et al., 1995, ApJ, 446,
L97

   \bibitem{cardelli88} Cardelli J.A., Clayton G.C., 1988, AJ, 95, 516

   \bibitem{witt80} Witt A.N., Cottrell 1980, A.J., 85, 22

   \bibitem{witt85} Witt A.N., 1985, ApJ, 294, 216

   \bibitem{witt86} Witt
A.N., Bohlin R.C, 1986, ApJ, 305, L23

   \bibitem{witt92} Witt A.N., et
al., 1992, ApJ, 395, L5

   \bibitem{zagury99} Zagury F., Boulanger F.,
Banchet V., 1999,
     A\&A, 352, 645


\end{thebibliography}
\end{document}